\documentclass[a4paper,12pt]{article}
\usepackage[margin=1in]{geometry}

\usepackage{hyperref}
\usepackage[utf8]{inputenc}
\usepackage{psfrag}
\usepackage{lineno}
\usepackage{amsmath,amssymb}
\usepackage{xcolor}
\usepackage{hyphenat}
\usepackage{comment}
\usepackage{stfloats}

\usepackage{graphicx}

\usepackage{ulem}

\usepackage{authblk}

\usepackage{algorithmic}

\title{On Key Epidemiological Metrics during Infectious Disease Outbreaks}
\author[1]{Hernán De~Battista\thanks{Corresponding author. Email address: deba@ing.unlp.edu.ar}}
\author[1]{Jos\'e Garc\'ia~Cl\'ua}
\author[1]{Sebastián Nu\~nez}
\author[1]{Fernando Inthamoussou}
\author[1]{Fabricio Garelli}
\affil[1]{Grupo de Control Aplicado GCA-LEICI. Facultad de Ingeniería, Universidad Nacional de La Plata -- CONICET, Argentina}
\date{October 26, 2020}

\begin{document}

\maketitle

\begin{abstract}
In this work, we review the figures used to characterize an epidemic outbreak most. Particular attention is drawn to epidemic spreading at time-varying transition rates. A time-varying SIR-like model is used to describe the epidemic dynamics, from which closed equations relating parameters and key quantities like reproduction ratio and doubling time are derived. The definition and computation of these metrics are revisited in the context of the general solution to the time-varying model dynamics, focusing on the similarities and differences with the time-invariant case. Further, the  prediction  of  these metrics, that is of the  disease  evolution, as response to different scenarios is also investigated. 
\end{abstract}



\section{Introduction}

Mathematical analysis and modeling help to better understand the transmission of infectious diseases  \cite{He:00}. They are essential to predict the speed of an emerging outbreak, as well as the peak of infection and total amount of infected people and deaths. These mathematical tools become crucial when trying to control unknown diseases having mortality and transmission potential to become epidemic.
Effectively, they can be used to identify potential public health interventions, to predict their impact and to retrospectively assess their efficacy \cite{Holl,mu2019optimal}. 

 This is the case of the COVID-19 disease first registered in China by December 2019 and rapidly spread with pandemic proportions and devastating effects \cite{Tagliazucchi2020}. Lockdown, social distancing, testing and tracing policies implemented by governments around the world have allowed tackling and mitigating the effects caused by the SARS-CoV-2 virus. These interventions have been usually decided on the basis of mathematical models derived from global scientific efforts \cite{FaYiMa:20,MoPaAmNo:20,metcalf2020mathematical}. 

During emerging outbreaks, disease spreading typically follows exponential growth dynamics \cite{ChSaBaVi:16,Ch:16}. The transmission process can be described by mass-action kinetics using differential equations resulting in analytic expressions \cite{Ba:75}. In this sort of models, the community members are aggregated into a few compartments identified with the different states of the disease. These models are characterized by the rates of transition among compartments. The model complexity is determined by the number of compartments and pathways required to describe the disease dynamics with sufficient accuracy \cite{He:00,KeKe:37}. An homogeneous population is implicitly assumed in the above definitions. Age, geographic, social-economic and activity heterogeneity can be included into the model yielding more compartments and a contact matrix connecting the groups  \cite{BoBoCaMa:18,huo2019dynamics,liu2019dynamics}. 

One fundamental concept in the study of infectious disease dynamics is the effective reproduction ratio $\mathcal{R}_e$. This metric is defined as the average number of secondary infections caused by a single individual during his infectious period. Based on this figure it can be predicted if, and how fast, the infection will spread \cite{He:05,Holl}. During the COVID-19 pandemic, this metric has been commonly used by authorities as a qualitative measure of disease evolution, with $\mathcal{R}_e<1$ being the objective pursued by official interventions \cite{lan2019stochastic}. 

Another metric used during an emerging outbreak is the doubling time $t_2$, which can generically be defined as the rate at which the number of people at each disease state doubles. The doubling time explicitly measures the speed of disease spreading. It can be used to predict ICU occupancy, diagnostic testing consumables requirement and other demands in the short term. Also, it can be employed to decide official interventions, with a given doubling time as the pursued objective, and to evaluate the efficacy of interventions \cite{reeves2020coronavirus}. Of course, differing from the reproduction ratio, the doubling time has only meaning before the outbreak peak \cite{KrSch:20}.

The doubling time can be calculated retrospectively from reported data. Although the outbreak is dominated by the infective individuals, the doubling time is often calculated using the data of cumulative infected people \cite{oliveiros2020role}. This is a common practice because the data of reported cases is more reliable than that of active infective people. This practice is sustained by the assumption that the number of individuals at different disease states increases in the same proportion during an exponential growth. On its side, the reproduction ratio can be computed in different ways from reported cases, tracing, testing and statistics. Both quantities, doubling time and reproduction ratio, can also be determined from the transition rates of mathematical models.

The aim of this work is to shed some light into the definitions, uses and differences among these metrics. For that purpose, a two-compartment model is preferred over  higher order models to gain in insight and closed mathematical expressions. The paper focuses on the transient dynamics of disease outbreaks, so transition rates are considered time-varying in the more general case.

The paper is structured as follows. Section \ref{sec:Models} presents the model used in the rest of the paper that captures the dominant dynamics in infectious diseases and introduces the most important transition rates.  Section \ref{sec:time-invariant-model} revisits the time-invariant exponential growth of a disease and, as a first contribution, derives some conditions for the validity of well-known relationships. Also, the solution from arbitrary initial conditions is used to quantify the discrepancies among different ways of calculating the doubling time and reproduction ratio. Section \ref{sec:time-varying-phase} addresses a more general time-varying scenario. In Section \ref{sec:prediction}, the prediction of the disease evolution is discussed with focus on the way an intervention can affect the involved metrics. Finally, general conclusions are drawn in Section \ref{sec:conclusions}.

\section{Epidemiological model} \label{sec:Models}

One of the simplest compartmental models in epidemiology, which is the basis for more complex ones is the SIR model \cite{He:00}. It aggregates the total population $N$ in three compartments: the susceptible $S$, the infectious $I$ and the removed $R$ ones. The equations describing the model dynamics are

\begin{align}
\begin{split}
\frac{dS(t)}{dt}  &= -\frac{\beta_0 I(t) S(t)}{N}, \\
\frac{dI(t)}{dt}  &= \frac{\beta_0 I(t) S(t)}{N} -\gamma I(t), \\
\frac{dR(t)}{dt}  &= \gamma I(t), \\
\end{split} 
\label{eq:SIR}
\end{align}
where $\beta_0$ is the average of contacts per person per time and $\gamma^{-1}$ is the average time period that an individual is infective. The infective individuals are capable of infecting susceptible ones. In fact, after an infectious contact, the susceptible individual transitions to the infectious compartment. Removed individuals are those that are no more infective, either because they recovered or died. 

Transition from compartment $S$ to compartment $I$ takes place at the rate $\frac{\beta_0 I(t) S(t)}{N}$ where $\frac{S(t)}{N}$ is the probability of contacting a susceptible individual. Then, $\frac{\beta_0 S(t)}{N}$ is the average of infections per time transmitted by an infective individual, called also infection rate or contagion rate. On the other hand, the transition from compartment $I$ to compartment $R$ occurs at the removal rate $\gamma$. 

The SIR model \eqref{eq:SIR} does not include the so-called vital dynamics, assuming births and deaths occur at much lower rates than the compartment transition rates. Otherwise, they should be included in the model. Also, \eqref{eq:SIR} can be modified to account for the loss of immunity, in which case removed individuals move back to the susceptible compartment at the immunity rate. This is the so-called SIRS model \cite{hu2019global}. Another variation of the SIR model is the SEIR one, which includes an intermediate compartment in the transition from $S$ to $I$. This compartment $E$ refers to exposed individuals who are infected but not yet infective. This model is used when the incubation period is comparable to the infectious one \cite{Tagliazucchi2020}.

The susceptible compartment has very important effects in the long-term evolution of the disease, steady state and stability. In contrast, when the disease is studied over a relatively short period of time, or the transition rates are derived from data, or during an initial exponential phase of a disease, the $S$ compartment can be disregarded. In that case, the epidemic can be described by a two-compartmental model. In this work, we conveniently rearrange the population in the infectious compartment $I$ and the cumulative infected compartment $C$. The $C$ compartment merges all individuals that have been infected, that is $C=I+R$:

\begin{align}
\begin{split}
\frac{dI(t)}{dt}  &= (\beta(t) - \gamma(t)) I(t), \\
\frac{dC(t)}{dt}  &= \beta(t) I(t).
\end{split} 
\label{eq:SIRC}
\end{align}

In \eqref{eq:SIRC}, the transition rates are explicitly written as function of time. The time-varying $\beta(t)$ accounts for the $S(t)/N$ ratio in \eqref{eq:SIR} as well as for changes in habits and interventions. For instance, the use of mouth caps or masks and hygiene care reduce the probability of a contact to be infectious, and therefore reduce $\beta$. Also $\beta$ is reduced by social distancing and lockdown. Obviously, $\beta$ increases back when these measures are relaxed. On the other side, the removal rate $\gamma$ varies with time because of new treatments, drugs, etc. Also, time-varying rates are typically obtained when fitting the model to collected data.


\section{Time-invariant exponential growth} \label{sec:time-invariant-model}

Consider the transition rates in \eqref{eq:SIRC} are constant:
\begin{align}
\begin{split}
\frac{dI(t)}{dt}  &= (\beta - \gamma) I(t), \\
\frac{dC(t)}{dt}  &= \beta I(t).
\end{split} 
\label{eq:SIRCc}
\end{align}
It is clear  that the model dynamics is governed by the infectives. In fact, the $I$ dynamics is autonomous, whereas the $C$ (and $R$) dynamics are given by integration of $I(t)$. We therefore focus for the moment on the $I$ dynamics. The solution to the first equation of \eqref{eq:SIRCc} is 

\begin{equation}
I(t)  = I(t_0) e^{((\beta - \gamma)(t-t_0))}.   
\label{eq:LTIsolI}
\end{equation}
The infectious compartment increases exponentially when the contagion rate is greater than the removal one ($\beta > \gamma$), decays exponentially when $\beta < \gamma$ and remains constant when $\beta = \gamma$. 

A measure commonly used to quantify the progress of a disease is the so-called reproduction ratio $\mathcal{R}_e$, which is defined as the average number of infections caused by a single infective individual during his infectious period \cite{He:05}. Taking into account that $\beta$ is the average infectious contacts per person per time and the average infectious period is  $\gamma^{-1}$, it is clear that the reproduction ratio can be written in terms of the transition rates of \eqref{eq:SIRCc} as
\begin{equation}
\mathcal{R}_e=\frac{\beta}{\gamma}.    
\label{eq:R0}
\end{equation}

Using \eqref{eq:R0}, the solution \eqref{eq:LTIsolI} can be rewritten as

\begin{equation}
I(t) = I(t_0) e^{(\gamma(\mathcal{R}_e - 1)(t-t_0))}. \\
\label{eq:LTIsolR0}
\end{equation}
Because of the definition of the reproduction ratio $\mathcal{R}_e$, it is of particular interest to determine the change in the infective population one infectious period $\gamma^{-1}$ ahead:
\begin{align}
\begin{split}
\frac{I(t+\gamma^{-1})}{I(t)} &= \frac{I(t)}{I(t-\gamma^{-1})}= e^{(\mathcal{R}_e-1)}, \\
\frac{I(t+\gamma^{-1})}{I(t)} &=\frac{I(t)}{I(t-\gamma^{-1})} \approx \mathcal{R}_e \quad \rm{if}\, \mathcal{R}_e\approx 1. \end{split} 
\label{eq:LTIgamma}
\end{align}
A reproduction ratio $\mathcal{R}_e>1$ means that the disease expands, whereas $\mathcal{R}_e<1$ means that the epidemic goes extinct. If $\mathcal{R}_e\approx 1$, the probability that an individual infects another one during the infection period of who infected him is very low, therefore the epidemic progresses in proportion to $\mathcal{R}_e$. Conversely, if $\mathcal{R}_e$ is high, this probability is not negligible and the epidemic grows faster than what $\mathcal{R}_e$ predicts. 
Obviously, in this time-invariant case, $\mathcal{R}_e$ applies forward and backward in time. The reproduction ratio uniquely determines the progress of a disease when the time unit is the infectious period. In fact, if time is scaled as
\begin{equation}
    d\tau=\gamma dt,
    \label{eq:tau}
\end{equation}
where $d\tau$ is the probability that an infective individual is removed from compartment $I$ during the time interval $dt$,
the disease dynamics is governed by 
\begin{align}
\begin{split}
  \frac{dI(\tau)}{d\tau}  &= (\mathcal{R}_e - 1) I(\tau), \\
  I(\tau)  &= I(\tau_0) e^{((\mathcal{R}_e - 1)(\tau-\tau_0))}, \\
  \end{split} 
\label{eq:SIRtau}
\end{align}
so that the infective people one average infectious period ahead is
\begin{equation}
\frac{I(\tau+1)}{I(\tau)} = \frac{I(\tau)}{I(\tau-1)} =e^{(\mathcal{R}_e-1)}.
\label{eq:SIRtau1}
\end{equation}

Another metric, closely related to the reproduction ratio, usually used to measure the progress of a disease is the doubling time $t_{2e}$. As its name suggests, it is the time elapsed until the infective population doubles.  The doubling time is easily obtained from \eqref{eq:LTIsolI}:

\begin{equation}
I(t+t_{2e}) = I(t) e^{((\beta - \gamma)t_{2e})} \equiv 2I(t), 
\label{eq:tdupIprevio}
\end{equation}
yielding 
\begin{equation}
t_{2e} = \frac{\ln(2)}{\beta - \gamma}.
\label{eq:tdupI}
\end{equation}

Obviously, the doubling time in a constant rate exponential phase is valid both forward and backward in time. Furthermore, in case the disease is in a contraction phase, the doubling time will take a negative value. 

By properties of logarithmic functions, $I(t)$ in  \eqref{eq:LTIsolI} can be rewritten as
\begin{equation}
I(t)=I(t_0) 2^{(t-t_0)/t_{2e}}.
\label{eq:I2}
\end{equation}

Both metrics $\mathcal{R}_e$ and $t_{2e}$ are obviously related by
\begin{align}
\begin{split}
 t_{2e} =& \frac{\ln(2)}{\gamma(\mathcal{R}_e-1)}, \\
\tau_{2e} =& \frac{\ln(2)}{\mathcal{R}_e-1}, 
\end{split} 
\label{eq:tdupIR0}
\end{align}
where $\tau_{2e}$ is the doubling time given in average infectious periods. For instance, for $\mathcal{R}_e= 1.69$, the amount of infective individuals doubles every average infectious period. 


\subsection{Permanent-state exponential phase}

The solution to \eqref{eq:SIRCc} from arbitrary initial conditions $I(t_0)>0$ and $C(t_0) \geq I(t_0)$ is
\begin{align}
\begin{split}
I(t)  &= I(t_0) e^{((\beta - \gamma)(t-t_0))}, \\
C(t)  &= C(t_0) + \frac{\beta}{\beta-\gamma}(I(t)-I(t_0)) \; {\rm if } \beta \neq \gamma, \\
C(t)  &= C(t_0) + \beta I(t_0) (t-t_0) \; {\rm if } \beta = \gamma.
\end{split} 
\label{eq:LTIsol1}
\end{align}

Let focus on the expanding case with $\beta>\gamma$, which means that $I$ grows exponentially and doubles every $t_{2e}$. Furthermore, consider initial conditions $I(t_0)$, $C^*(t_0)=\dfrac{\beta}{\beta-\gamma} I(t_0)$. The particular solution from them is the following permanent-state exponential solution 
\begin{align}
\begin{split}
I(t)  &= I(t_0) 2^{(t-t_0)/t_{2e}}, \\
C(t)  &= C^*(t_0)  2^{(t-t_0)/t_{2e}},
\end{split} 
\label{eq:LTIsol}
\end{align}
which means that $C$ grows in proportion to $I$ from $t_0$: 
\begin{equation}
C(t) = \frac{\beta}{\beta - \gamma} I(t) = \frac{\mathcal{R}_e}{\mathcal{R}_e - 1} I(t).
\label{eq:Cprop}
\end{equation}
That is $C^*(t_0)$ is compatible with the permanent-state exponential growth. 

Because of the proportionality among variables, equations \eqref{eq:LTIgamma}, \eqref{eq:SIRtau1}, \eqref{eq:tdupI} and \eqref{eq:tdupIR0} apply also to $C(t)$. This also means that the doubling time can be indistinctly calculated from the data of infective or cumulative infected population.

It should be noticed that a permanent-state exponential decay cannot be defined. In fact, from \eqref{eq:SIRCc}, $C$ is always increasing, whereas $I$ is decreasing whenever $\beta<\gamma$. So, for $C(t)$ to be proportional to $I(t)$ the condition $C(t)<0$ would be required, which is obviously infeasible.


\subsection{Exponential phase from arbitrary initial conditions} \label{sec:arbitrary-initial-conditions}

Reconsider the solution to \eqref{eq:SIRCc} from arbitrary initial conditions $I(t_0)>0$ and $C(t_0) \geq I(t_0)$. Comparing with the particular solution \eqref{eq:LTIsol}, during an exponential growth \eqref{eq:LTIsol1} can be rewritten as
\begin{align}
\begin{split}
I(t)  &=I(t_0) 2^{(t-t_0)/t_{2e}}, \\
C(t)  &= \frac{\beta}{\beta - \gamma} I(t) ( 1- \delta(t) ), \\
\delta(t)  &= \delta(t_0) 2^{-(t-t_0)/t_{2e}},  \\
\delta(t_0) &= 1-\frac{C(t_0)}{C^*(t_0)}.
\end{split} 
\label{eq:LTIsol2}
\end{align}
The dynamics of $I$ remains the same, and so the doubling time \eqref{eq:tdupI}. Now, $C(t)$ is not proportional to $I(t)$ because of $\delta(t)$ that vanishes exponentially. Depending on the initial error $\delta(t_0)$, it may take several doubling times for $\delta(t)$ to be negligible. More precisely, for any $0<a<1$ and $\delta>a$, $|\delta(t)|<a$ only if $t-t_0>t_{2e} \log_2\left(\frac{\delta(t_0)}{a} \right)$. That is, it may take several doubling times for the cumulative cases $C$ to approach the permanent-state exponential growth. 

This transient behavior in $C$-dynamics has some important implications. Particularly, the doubling time for infective and cumulative infected populations are no more the same even though the transition rates are constant. We can say that the disease reaches the new permanent-state exponential phase just when $\delta(t)$ vanishes. During the transient, although transition rates are constant and doubling time of infective people is constant, the doubling time of cumulative cases $t_{2C}$ varies with time. Indeed, for any $t>t_0$, the doubling time $t_{2C}$ satisfies $C(t+t_{2C})=2C(t)$: 
\begin{align}
\begin{split}
I(t+t_{2C}) ( 1- \delta(t+t_{2C}) ) &= 2 I(t) (1-\delta(t)), \\
 I(t)2^{\frac{t_{2C}}{t_{2e}}} - I(t) \delta(t)  &= 2 I(t) (1-\delta(t)), \\
2^{\frac{t_{2C}}{t_{2e}}} - \delta(t) &= 2 (1-\delta(t)), \\
\end{split} 
\label{eq:T2caux}
\end{align}
thus,
\begin{equation}
t_{2C}(t) = t_{2e} \log_2 (2-\delta(t)) = \frac{\ln(2-\delta(t))}{\beta-\gamma}.
\label{eq:T2c}
\end{equation}
It is common practice to compute the doubling time from confirmed cases data, in which case the  relative error $e_{t_{2C}}$ between $t_{2C}(t)$ and $t_{2e}$ is obtained by comparison with \eqref{eq:tdupI}:
\begin{equation}
e_{t_{2C}}(t) \equiv \frac{t_{2C}-t_{2e}}{t_{2e}}=\log_2 \left( 1 - \frac{\delta(t)}{2}\right).    
\end{equation}

The time $t_\epsilon$ elapsed to bound the error $|e_{t_{2C}}(t_\epsilon+t_0)| < \epsilon$ provided $|e_{t_{2C}}(t_0)| > \epsilon$ can be derived from the previous equations:
\begin{equation}
t_\epsilon=t_{2e} \log_2\frac{\delta(t_0)}{2(1-e^{\epsilon sign(\delta(t_0))})}.
\label{eq:error}
\end{equation}

Also, the relationship among the reproduction of $C$ and the reproduction ratio $\mathcal{R}_e$  can be similarly determined. In fact, from the second equation of \eqref{eq:LTIsol2} evaluated at $t$ and one infectious period ahead yields
\begin{equation}
\frac{C(t+\gamma^{-1})}{C(t)} = e^{(\mathcal{R}_e-1)}  \frac{1-\delta(t)e^{(1-\mathcal{R}_e)}}{1-\delta(t)}.
\label{eq:LTIgamma1}
\end{equation}
As expected, the reproduction of $C$ approaches that of the permanent-state exponential growth as $\delta(t)$ vanishes.


\subsection{Example~1}

This example aims to illustrate the evolution of the doubling time during the initial transient of an exponential phase from arbitrary initial conditions. Furthermore, the error of approximating the doubling time using data of cumulative infected cases $C$ is quantified. Suppose the disease grows exponentially until $t=0$ with a constant reproduction ratio $\mathcal{R}_e^-$. At $t=0$, the contagious rate $\beta$ is stepped so that a new exponential phase with reproduction ratio $\mathcal{R}_e^+$ starts from initial conditions determined by the previous exponential phase. Two cases are considered:
\begin{enumerate}
    \item $\mathcal{R}_e^-=1.5$ and $\mathcal{R}_e^+=2$.
    \item $\mathcal{R}_e^-=3$ and $\mathcal{R}_e^+=2$.
\end{enumerate}    
The doubling time for $t \geq 0$ (measured in infectious periods)  is $\tau_{2e}^+=\ln(2)\approx 0.69$. The initial value of $\delta(t)$ can be computed from the reproduction ratios before and after $t=0$:
\begin{align}
\begin{split}
C(0)&= \frac{\mathcal{R}_e^-}{\mathcal{R}_e^- - 1} I(0),\\
C^*(0)&= \frac{\mathcal{R}_e^+}{\mathcal{R}_e^+ - 1} I(0), \\
\delta(0) &= \frac{\frac{\mathcal{R}_e^-}{\mathcal{R}_e^+}-1}{\mathcal{R}_e^- - 1},
\end{split} 
\label{eq:ex1}
\end{align}
so that $e_{t_{2C}}(0)= \log_2(5/4)\approx 0.32$ and  $e_{t_{2C}}(0)= \log_2(7/8)\approx -0.19$ for the two cases, respectively.

\begin{figure}
 \centering
 \includegraphics[width=0.95\columnwidth]{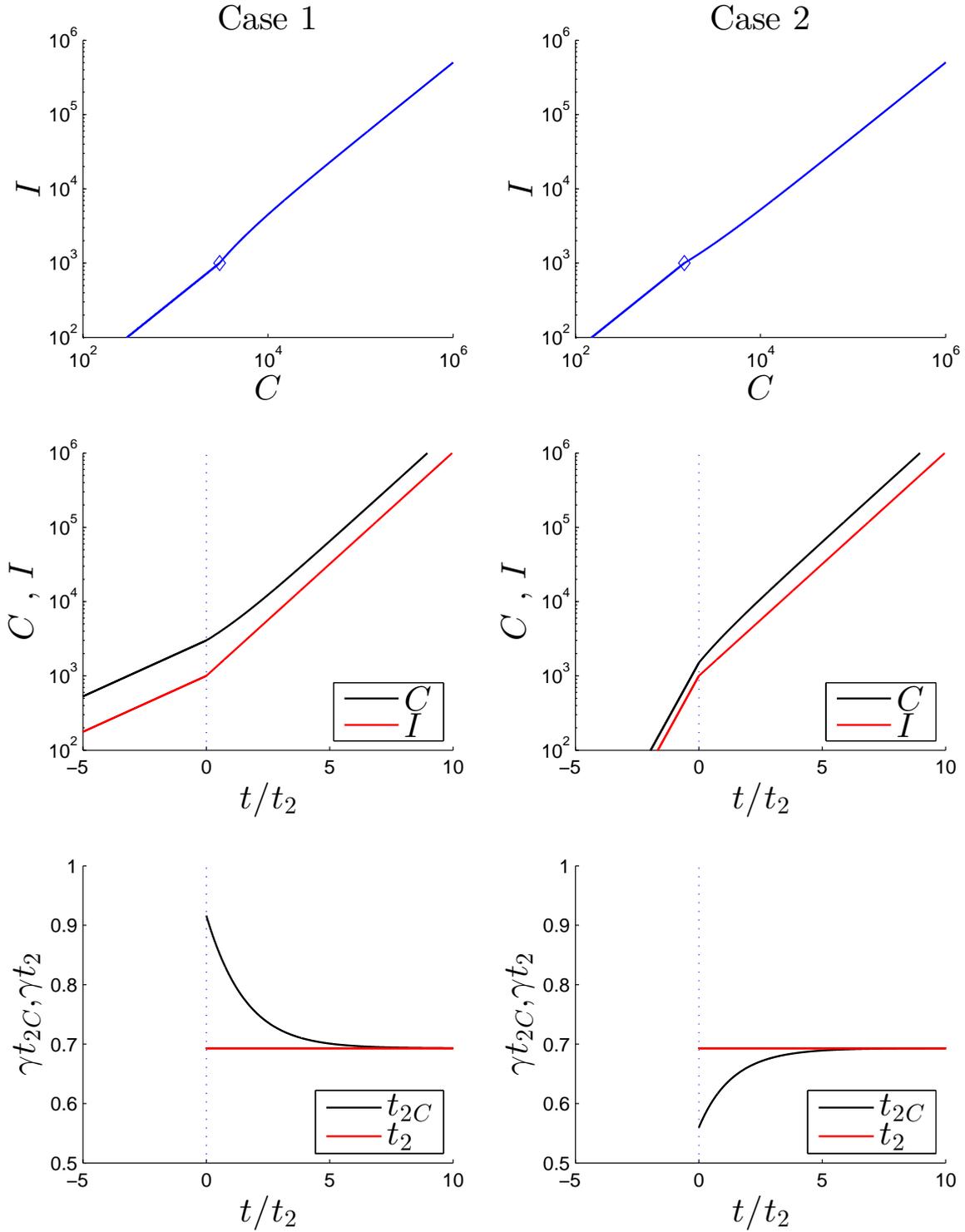}
  \caption{Doubling times for exponential phase with arbitrary initial conditions. Left and right columns: cases 1 and 2, respectively. Top row: state planes $C - I$. Middle row: time response of both  cumulative infected (black) and infective (red) populations. Bottom row: doubling time for cumulative infected ($t_{2C}$ in black) and infective ($t_2=t_{2e}^+$ in red) population.}
 \label{fig:t2c}
 \end{figure}

Figure~\ref{fig:t2c} shows the evolution of the doubling times during the initial transient of the exponential phases from arbitrary initial conditions. The left and right columns correspond to cases 1 and 2, respectively. The top row depicts the state planes $C - I$, with a diamond indicating time $t=0$. The middle row shows the time response of both  cumulative infected (black) and infective (red) populations. It can be observed that the condition at $t=0$ is imposed by the previous exponential phase. Finally, the bottom row compares the doubling time for cumulative infected ($t_{2C}$ in black) and infective ($t_2=t_{2e}^+$ in red) population. The doubling time is $t_{2e}^+=\gamma^{-1} \ln(2)$ for all $t>0$, while the doubling time of cumulative infected population $t_{2C}$ varies with time and converges to $t_{2e}^+$ after 5 doubling times approx. Therefore, we can say that it takes 5 doubling times for the disease to reach the new exponential phase. However, a shorter time would be probably needed to bound the error $|e_{t_{2C}}|$. For instance, if an error $\epsilon=.05$ is acceptable, a period $t_\epsilon \approx 2.8 t_{2e}^+$ and $t_\epsilon \approx 1.9 t_{2e}^+$ would be enough according to \eqref{eq:error}, respectively. Anyway, the transient form arbitrary initial conditions to the permanent-state exponential growth is longer as  the disease growth is slower.

\section{Time-varying exponential growth} \label{sec:time-varying-phase}

Let go back to the time-varying model \eqref{eq:SIRC}. The solution of the differential equation for $I$ is:
\begin{equation}
I(t) = I(t_0) e^{\left(\int_{t_0}^t (\beta(s)- \gamma(s))ds\right)}.
\label{eq:Itv}
\end{equation}
The doubling time $t_{2}(t)$ obviously varies with time, and can be defined forward ($t_{2}^{+}(t)$) and backward ($t_{2}^{-}(t)$) in time:
\begin{align}
\begin{split}
I(t+t_{2}^{+}(t)) &\equiv 2I(t),  \\
I(t) &\equiv 2 I(t-t_{2}^{-}(t)).  
\end{split} 
\label{eq:tdupI+-}
\end{align}
From \eqref{eq:Itv}--\eqref{eq:tdupI+-}, $t_{2}^{+}(t)$ y $t_{2}^{-}(t)$ are the solutions of
\begin{align}
\begin{split}
\int_{0}^{t_{2}^{+}} (\beta(t+s)- \gamma(t+s))ds &= \ln(2), \\
\int_{0}^{t_{2}^{-}} (\beta(t-s)- \gamma(t-s))ds &= \ln(2).               
\end{split} 
\label{eq:tduptv}
\end{align}

\begin{figure}
 \centering
 \includegraphics[bb= 1 150 600 700,width=0.95\columnwidth]{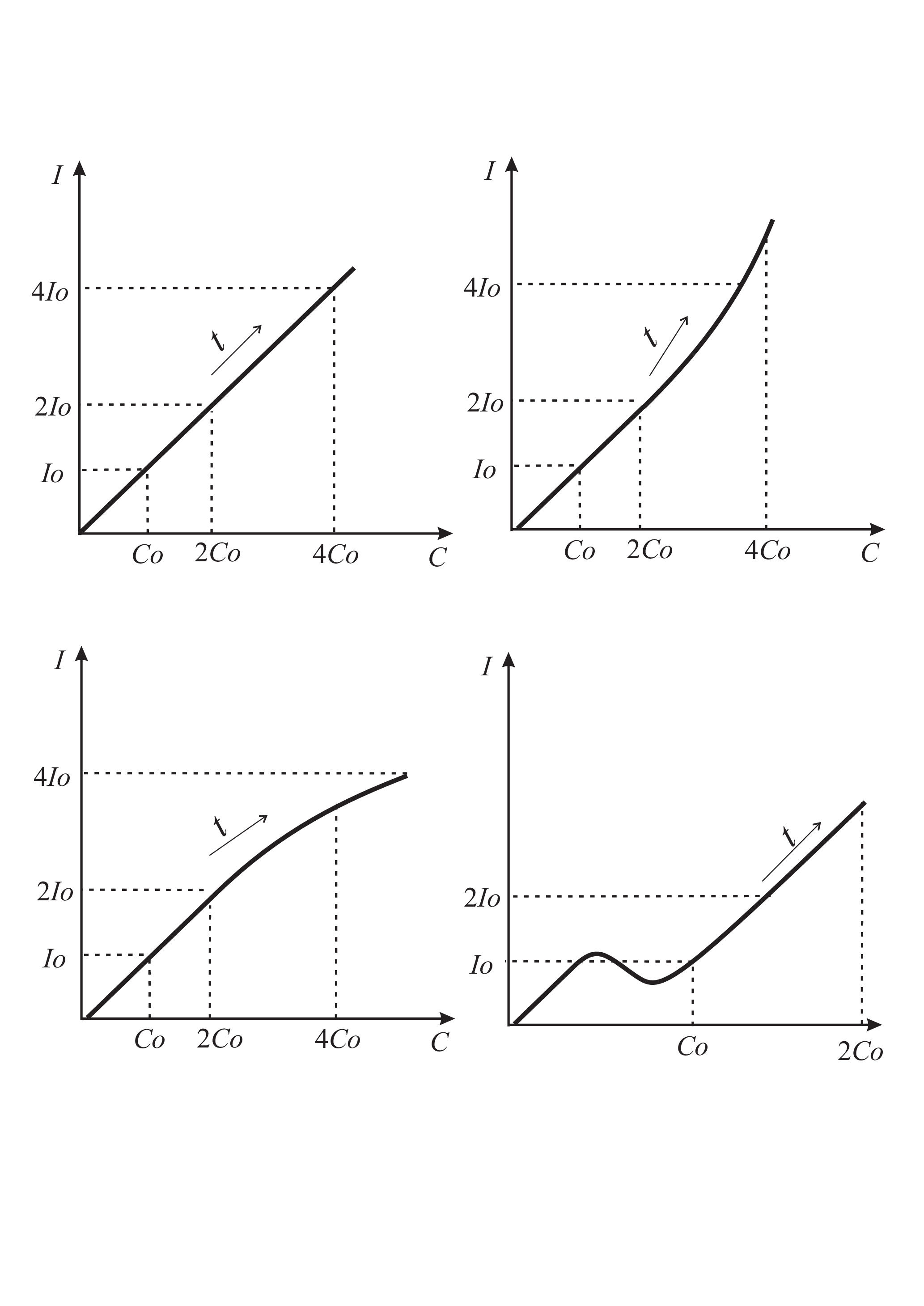}
  \caption{Doubling times for time-varying disease growth. Top-left plot: constant-rate exponential phase. Bottom-left plot: disease deceleration. Top-right plot: disease in acceleration. Bottom-right plot: regrowth of the disease from ($I_0 , C_0$).}
 \label{fig:t2tv}
 \end{figure}

The computation of the doubling time for the $C$-population ($t_{2C}(t)$) requires integrating the second equation of \eqref{eq:SIRC} after replacing $I(t)$ with \eqref{eq:Itv}. As can be expected, $t_{2C}(t)$ generally differs from $t_{2}(t)$. Figure~\ref{fig:t2tv} qualitatively compares these doubling times. 

 The top-left plot illustrates a constant-rate exponential phase where doubling of $C$ and $I$ occurs simultaneously ($t_{2C}=t_2$). The bottom-left plot depicts the case of disease deceleration from ($2I_0 , 2C_0$) with infective population growing slower than the cumulative infected one, thus resulting in $t_{2C}(t)<t_2(t)$. The case of a disease in acceleration is shown in the top-right plot, with the infective population growing faster from ($2I_0 , 2C_0$), so that $t_{2C}(t)>t_2(t)$. Finally, the bottom-right plot illustrates the case of a regrowth of the disease from ($I_0 , C_0$), where the previous conclusion is much more evident. In many cases, particularly in the latter one, the initial conditions should be reset to reduce the difference among doubling times.

In the same way as with doubling time, the reproduction ratio can be defined forward ($\mathcal{R}^+(t)$) and backward ($\mathcal{R}^-(t)$) in time. It is convenient to scale time as in \eqref{eq:tau}:
\begin{equation}
    \tau=\int \gamma(t) dt,
    \label{eq:tautv}
\end{equation}
so that
\begin{equation}
  \frac{dI(\tau)}{d\tau}  = \left(\frac{\beta(\tau)}{\gamma(\tau)} - 1\right) I(\tau).
\label{eq:SIRtautv}
\end{equation}

Defining
\begin{align}
\begin{split}
\mathcal{R}^+ (\tau) &=\int_{0}^{1} \frac{\beta(\tau+s)}{\gamma(\tau+s)} ds,  \\
\mathcal{R}^- (\tau) &=\int_{0}^{1} \frac{\beta(\tau-s)}{\gamma(\tau-s)} ds, 
\end{split} 
\label{eq:Rtv}
\end{align}
yields expressions similar to \eqref{eq:SIRtau1}:
\begin{align}
\begin{split}
\frac{I(\tau+1)}{I(\tau)} &= e^{(\mathcal{R}^+ (\tau)-1)}, \\
\frac{I(\tau)} {I(\tau-1)} &= e^{(\mathcal{R}^-(\tau)-1)}.
\end{split} 
\label{eq:SIRtautv2}
\end{align}

Obviously, since they are defined over different time windows, there is no more a direct relationship between doubling time and reproduction ratio as in the time invariant case (see \eqref{eq:tdupIR0}). Solution for $t_2^\pm(t)$ to \eqref{eq:tduptv} can be obtained numerically in the general case. A possible algorithm that makes use of a numerical integrator (e.g. a trapezoidal method) is described in Fig. \ref{algoritmo.eq27}.

\begin{figure}
\begin{algorithmic}
\REQUIRE vectors $t,\beta(t),\gamma(t)$ and time instant $t^*$.
\STATE $t_0 \leftarrow t(1)$
\STATE $t_{f}\leftarrow t(end)$
\STATE $F_a \leftarrow$ CumulativeIntegral($\beta-\gamma,t,[t^*,t_f]$)
\STATE $R_a \leftarrow$ CumulativeIntegral($\beta-\gamma,t,[t_0,t^*]$)
\STATE MaxFa $\leftarrow$ max of ($F_a$)
\STATE MaxRa $\leftarrow$ max of ($R_a$)

\STATE $idx_{t^*} \leftarrow$ index of vector t corresponding to $t^*$

\IF {MaxFa $<$ ln(2)} 
\STATE $t_2^+ \leftarrow$ NaN
\ELSE
\STATE $idx_{min} \leftarrow$ index of min(abs($F_a-$ln(2)))
\STATE $idx_{t_2^+} \leftarrow idx_{t^*} + idx_{min}$
\STATE $t_2^+ \leftarrow t(idx_{t_2^+}) - t^*$
\ENDIF

\IF {MaxRa $<$ ln(2)} 
\STATE $t_2^- \leftarrow$ NaN
\ELSE
\STATE $idx_{min} \leftarrow $ index of min(abs($R_{a(end)}-R_a-$ln(2)))
\STATE $idx_{t_2^-} \leftarrow idx_{min}$
\STATE $t_2^- \leftarrow t^* - t(idx_{t_2^-})$
\ENDIF 

\RETURN{$t_2^+,t_2^-$}
\end{algorithmic}
\caption{Numerical algorithm for solving $t_2^+$, $t_2^-$ in eq. \eqref{eq:tduptv} for the function $z(t)=\beta(t)-\gamma(t)$ at $t = t^*$. Function CumulativeIntegral($\cdot$) returns a vector corresponding to the cumulative integral at each sampling time.\label{algoritmo.eq27}}
\end{figure}


\subsection{Example~2}
\begin{figure}
\centering
\includegraphics[width=0.85\columnwidth]{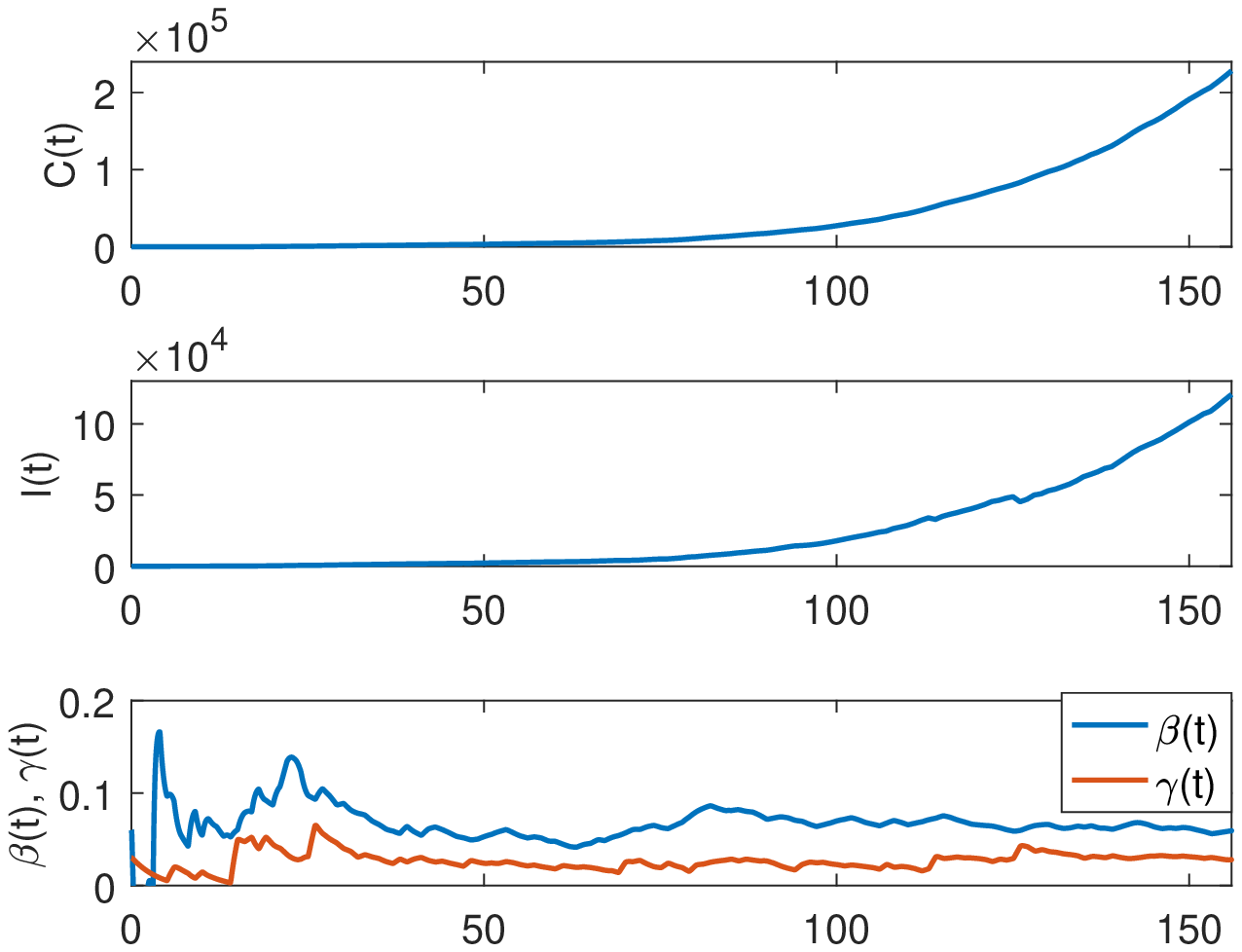}
\includegraphics[bb= 0 108 420 305, width=0.85\columnwidth]{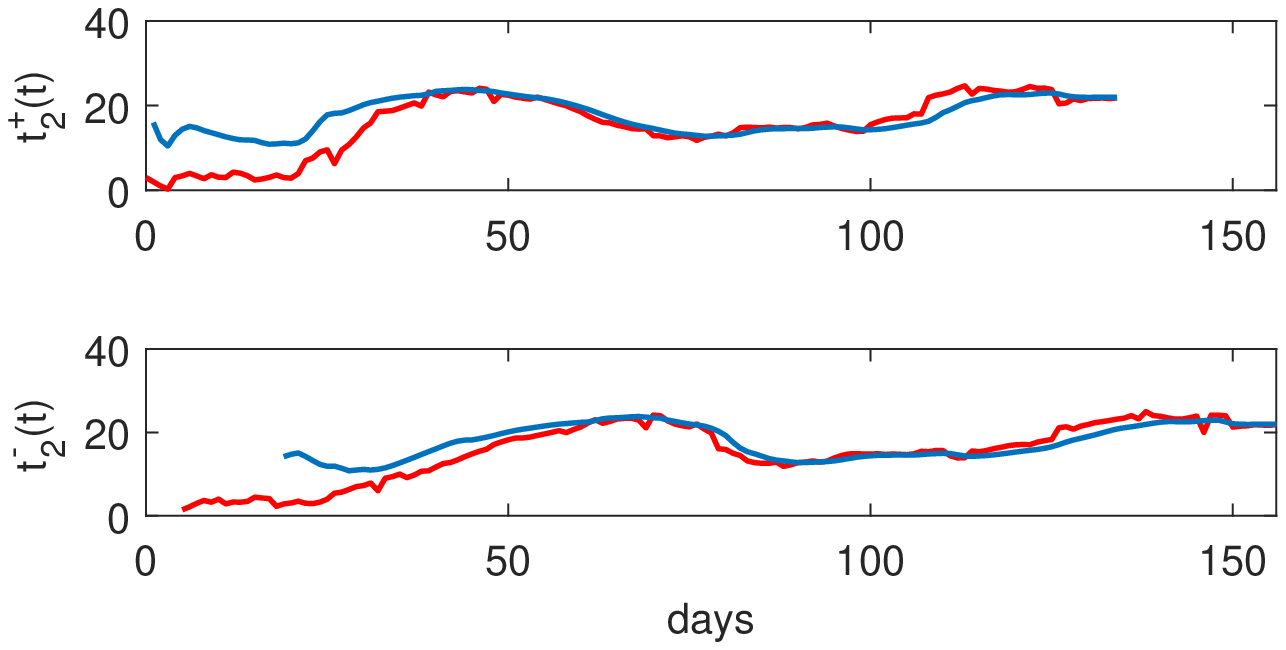}
\caption{Reconstruction of $t_2^+(t)$ and $t_2^-(t)$ by solving eq. \eqref{eq:tduptv} with the algorithm shown in Fig. \ref{algoritmo.eq27} (blue lines) compared with direct calculation from data (red lines). \label{fig:example2}}
\end{figure}

Figure \ref{fig:example2} shows the reconstruction of $t_2^+$, $t_2^-$ based on the algorithm described in Fig. \ref{algoritmo.eq27}. The input data correspond to Argentina COVID-19 cases from March 3rd to August 6th (157 days). From top to bottom the plotted signals are $C(t)$, $I(t)$, $\widehat{\beta}(t)$ and $\widehat{\gamma}(t)$, $t_2^+(t)$ and $t_2^-(t)$. Time-varying estimates of $\beta$ and $\gamma$ denoted as $\widehat{\beta}(t)$, $\widehat{\gamma}(t)$ were obtained with an observer. The convergence errors of the observer are responsible for the discrepancies among the different computations of the doubling times.

\section{Prediction} \label{sec:prediction}

Computing $t_2^+(t)$ requires knowing the transition rates for a time window $[t,t+t_2^+(t)]$, where $t_2^+(t)$ is to be calculated. Therefore, when $t_2^+(t)$ is being computed on-line, a prediction of the transition rates $\hat\beta$ and $\hat\gamma$ for a long enough time horizon is required.


\subsection{Zero-order prediction}

The simplest prediction is 

\begin{align}
\begin{split}
\hat\beta(t+s) &=\beta(t) \; \forall s>0,\\
\hat\gamma(t+s) &=\gamma(t) \; \forall s>0.
\end{split} 
\label{eq:ratesctes}
\end{align}

In practice, this is equivalent to assume an exponential phase from arbitrary initial conditions begins at $t$ with constant transition rates $\beta(t)$ and $\gamma(t)$. The estimated doubling time of the infective population $\hat{t}_2^+(t)$ is, as expected, 
\begin{equation}
 \hat{t}_2^+(t)=t_{2e}(t)= \frac{\ln(2)}{\beta(t)-\gamma(t)}.
\label{eq:t2etv}
\end{equation}

We have previously studied the discrepancy among doubling times for cumulative infected and active infective populations under exponential phases from arbitrary initial conditions.


\subsection{First-order prediction}

A prediction that accounts for the drift of the transition rates can alternatively be used to compute $t_2^+$ from \eqref{eq:tduptv}. Consider, 

\begin{align}
\begin{split}
\hat\beta(t+s) &=\beta(t)+\dot\beta(t) s \; \forall s>0,\\
\hat\gamma(t+s) &=\gamma(t) \; \forall s>0,
\end{split} 
\label{eq:rateslinear}
\end{align}
where, for the sake of brevity and without loss of generality, a zero-order prediction of $\gamma(t+s)$ is used.

The estimated doubling time $\hat{t}_2^+(t)$ can be derived from \eqref{eq:tduptv}. After some trivial algebraic steps, one arrives to the following expression as function of  $t_{2e}(t)$:
\begin{equation}
 \hat{t}_2^+(t) = \frac{\ln(2)}{\dot\beta(t)t_{2e}(t)} \left( \sqrt{1+2\frac{\dot\beta(t)}{\ln(2)}t_{2e}(t)}-1\right),
\label{eq:t2tv}
\end{equation}
that for small enough drift $\dot\beta(t)$ approximates 
\begin{equation}
 \hat{t}_2^+(t) = t_{2e}(t) \left(1 - \frac{\dot\beta(t) t_{2e}^2(t)}{\ln(4)} \right).
\label{eq:t2tvapprpx}
\end{equation}

In the case that a first-order prediction $\gamma(t+s)=\gamma(t) + \dot\gamma(t) s$ is used, then $\dot\beta(t)$ in \eqref{eq:t2tv}--\eqref{eq:t2tvapprpx} should be replaced with $(\dot\beta(t)-\dot\gamma(t))$.


\subsection{Model-response prediction}

Consider that some intervention is taken in order to set a new contagion rate, and that the response of the contagion rate to this intervention is known beforehand. The transient from the state before intervention to the new steady-state is modeled here in the following general form
\begin{align}
\begin{split}
\beta(t)  &= \beta^+ - (\beta^+ - \beta^-) f(t), \\
  \gamma(t) &= \gamma,
\end{split} 
\label{eq:betatv}
\end{align}
where $\beta^- =  \beta( t\leq 0)$, $\beta^+ =  \beta( t\geq T)$,  $f(t)$ is continuous and non-increasing function satisfying

\begin{align}
\begin{split}
   f(t)  & \equiv 1 \;  \forall \, t \leq 0,     \\
   f(t)  & > 0 \;  \forall \, t < T,     \\
   f(t)  & \equiv 0 \;  \forall \, t \geq T.     
\end{split} 
\label{eq:f}
\end{align}

We are particularly interested in quantifying the effect of the transient on the doubling time. So, we consider the case 
$0 \leq t<T<t+t_2(t)$. 
Defining $\delta\beta=\beta^+ - \beta^-$, Equation \eqref{eq:tduptv} becomes
\begin{align}
\begin{split}
& \int_{0}^{\hat{t}_{2}^+} (\beta(t+s) - \gamma(t+s)) ds = \\
&= \int_{0}^{\hat{t}_{2}^{+}} (\beta^+ - \gamma)ds - \int_{0}^{T-t}   \delta\beta f(t+s) ds  = \ln(2) \\
&= (\beta^+ - \gamma)\hat{t}_{2}^+ -  \delta\beta (F(T)-F(t))  = \ln(2),
\end{split} 
\label{eq:tduppredtv}
\end{align}
where $F(\cdot)$ is the primitive of $f(\cdot)$. Using the mean value theorem, for each $t\in [0,T]$, $\exists t^*\in(t,T)$ such that $F(T)-F(t)=f(t^*)(T-t)$ so that

\begin{equation}
  \hat{t}_2^+(t) = t_{2e}^+ \left(1 + \frac{f^*(t)}{\ln(2)} \delta\beta (T-t)  \right),
 \label{eq:tduppredtv2}
\end{equation}
where $t_{2e}^+=\frac{\ln(2)}{\beta^+ - \gamma}$ is the doubling time of the constant-rates exponential phase starting at $t=T$ and $0<f^*(\cdot)<1$. Particularly, at $t=0$:
\begin{equation}
  \hat{t}_2^+(0) = t_{2e}^+ \left(1 + \frac{\delta\beta}{\ln(2)} F(T)  \right),
 \label{eq:tduppredtv20}
\end{equation}
with $F(T)<T$. Therefore, the error of assuming that the doubling time after $t=0$ is that of the ongoing exponential phase, ignoring the transient settled by $f(t)$ is
\begin{align}
\begin{split}
 e_{t_2}(t) & =\frac{\hat{t}_2^+(t)-t_{2e}^+}{t_{2e}^+} = \frac{\delta\beta}{\ln(2)} f^*(t) (T-t), \\
 e_{t_2}(0) & =\frac{\hat{t}_2^+(0)-t_{2e}^+}{t_{2e}^+} = \frac{\delta\beta}{\ln(2)} F(T). \\
 \end{split} 
\end{align}
Since $0<f^*(\cdot)<1$, an upper-bound for $e_{t_2}(t)$ is  $e_{t_2}(t) < E_{t_2}(t)=\frac{\delta\beta}{\ln(2)} (T-t)$. In the previous equations, the step $\delta\beta$ can be written in terms of the doubling time  (or reproduction ratio) of the past and following exponential phases $t_{2e}^-=\frac{\ln(2)}{\beta^- - \gamma}$ ($\mathcal{R}e^-=\frac{\beta^-}{\gamma}$) and $t_{2e}^+=\frac{\ln(2)}{\beta^+ - \gamma}$ ($\mathcal{R}e^+=\frac{\beta^+}{\gamma}$):
\begin{equation}
 \delta\beta = \ln(2) \left( (t_{2e}^+)^{-1} - (t_{2e}^-)^{-1} \right)=\gamma (\mathcal{R}e^+ - \mathcal{R}e^-)
\label{eq:betatv2}
\end{equation}


\subsection{Example 3}

Consider a piece-wise linear contagion rate such that $\mathcal{R}_e^-=2$, $\mathcal{R}_e^+=1.2$, $f(t)=1-t/T$, with $T=\frac{1}{2}\gamma^{-1}$. The primitive of $f(t)$ is $F(t)=t \left(1-\frac{t}{2T}\right)$, so $F(T)=T/2$. The doubling time for the previous and next exponential phases are, in infection period units, $\tau_{2e}^-=0.69$ and $\tau_{2e}^+=3.47$, while the transient duration is $\mathcal{T}=\gamma T=1/2$. Therefore, evaluating \eqref{eq:tduppredtv20}--\eqref{eq:betatv2} at the above values renders $\hat\tau_{2}(0)=\tau_{2e}^+(1+e_{t_2}(0)) = 2.47$ with a relative error w.r.t. $t_{2e}^+$ equal to $e_{t_2}(0)=\frac{\mathcal{R}e^+ - \mathcal{R}e^-}{4\ln(2)}= -0.29$. That is, because of the linear transient, the doubling time at $t=0$ is about $30\%$ shorter than the expected one for the new exponential phase.


\section{Conclusions} \label{sec:conclusions}

The definitions of key metrics that characterize the strength of infectious disease outbreaks as function of transition rates were generalized to cope with time-varying exponential growths. It was shown that, even for the case of constant growth rate, the differences in calculating these metrics from active and confirmed infections can be large and vanish too slow. Thus, their computation using confirmed cases should be discouraged despite being more reliable data. The results allow also predicting these metrics from estimated time-varying transition rates. These results can be used to determine how much and how fast the contagion rate must be reduced in order to achieve a desired objective. Therefore, they provide valuable tools for epidemic outbreaks control.


\section*{Acknowledgements}
This work was supported by CONICET PIP 2015-0837, ANPCyT PICT 2017-3211, UNLP 2020-I253 and MINCYT BSAS28 COVID FEDERAL.

\normalem

\end{document}